\documentstyle{mn}
\title[High-resolution imaging of faint blue galaxies]
{High-resolution imaging of faint blue galaxies}
\author[M.M.Colless \etal]{
Matthew Colless,$^1$\thanks{Visiting Astronomer, Canada-France-Hawaii
Telescope, operated by the National Research Council of Canada, the
Centre National de la Recherche Scientifique de France and the
University of Hawaii}
David Schade,$^{2{\textstyle\star}}$
T.J.Broadhurst$^3$ and R.S.Ellis$^2$ \\
$^1$Mount Stromlo Observatory, Australian National University,
    Weston Creek, ACT 2611, Australia \\
$^2$Institute of Astronomy, Madingley Road, Cambridge CB3 0HA \\
$^3$Dept of Physics and Astronomy, Johns Hopkins University, Baltimore,
MD~21218, USA}
\date{Accepted 1994 January 7. Received 1993 October 13; in original
form 1993 October 13}

\def\etal{\hbox{et~al.}}
\def\ie{\hbox{i.e.}}
\def\eg{\hbox{e.g.}}

\def\kpc{\hbox{~h$^{-1}$~kpc}}
\def\EW{\hbox{W$_\lambda$}}
\def\la{\mathrel{\hbox{\rlap{\hbox{\lower4pt\hbox{$\sim$}}}\hbox{$<$}}}}
\def\ga{\mathrel{\hbox{\rlap{\hbox{\lower4pt\hbox{$\sim$}}}\hbox{$>$}}}}

\def\arcsec{\hbox{$^{\prime\prime}$}}

\begin{document}
\maketitle

\begin{abstract}
We have used HRCam on the CFHT to obtain subarcsecond images of 26
galaxies with z=0.1--0.7 from the redshift survey of Colless \etal. The
primary sample of 17 galaxies have enhanced star formation indicated by
[OII] equivalent widths greater than 20\AA, while the comparison sample
of 9 galaxies have equivalent widths less than 10\AA. By fitting
exponential disks or $r^{1/4}$ bulges to B, V and I images we have
derived scalelengths for the blue and red stellar populations and so
established the location of the star-formation (in the nucleus or the
disk) for each galaxy. We have also searched for nearby faint
companions in order to determine whether the star-formation might be
linked to tidal interactions or mergers. We find that these
moderate-redshift galaxies generally have straightforward low-redshift
analogues, in that their colours, sizes and luminosities are consistent
with those of various types of z$\approx$0 galaxies. The star-forming
objects have structural components consistent with the full range of
present-day disk galaxies, and absolute magnitudes spanning the range
M$^\ast$$-$1 to M$^\ast$+5. Some of these galaxies have star-formation
concentrated in their nuclei but most have star-formation occurring
across the entire disk. We find companions at projected distances
closer than 10\kpc\ for 30\% of the galaxies with enhanced
star-formation, whereas none of the comparison sample have such close
companions. This fraction is very similar to the 40\% excess in the
number of star-forming galaxies found in the redshift survey of Colless
\etal, and provides the first direct evidence linking interactions or
mergers to the increased fraction of field galaxies with enhanced
star-formation at moderate redshifts.
\end{abstract}
\begin{keywords}
galaxies: evolution -- galaxies: formation -- galaxies: interactions --
galaxies: starburst.
\end{keywords}

\section{INTRODUCTION}

At B$\approx$22 there are approximately twice as many galaxies in the
number counts as would be expected if there had been no evolution of the
galaxy population. Statistically-complete redshift surveys to B=22.5
(Broadhurst \etal\ 1988, Colless \etal\ 1990, 1993) have demonstrated
that the bulk of this excess comprises galaxies at modest redshifts
(z$<$0.5) rather than luminous young star-forming galaxies at
z$\sim$2--3 as was once supposed. More recent redshift surveys (Cowie
\etal\ 1991, Glazebrook \etal\ 1993) have extended this result to B=24,
finding a median redshift of only z$\approx$0.4 and few galaxies with
z$>$1 even though the number counts exceed the no-evolution model by a
factor of five at this depth.

\begin{table*}
\centering
\caption{Sample of galaxies.}
\begin{tabular}{ccccccccl}
\multicolumn{9}{l}{(a) 1992 April 7--8} \\
ID&\multicolumn{2}{c}{R.A.~~~~(1950)~~~~Dec.}&b$_J$&b$_J$$-$r$_F$&
z&\EW(\AA)&~M$_{b_J}$&V (A,B,C)\\[10pt]
10.2.05&10 44 02.86&+00 00 58.5&21.72&1.91&0.303&~4&$-$19.2&20.7\\
10.2.17&10 44 04.53&+00 00 18.3&21.75&1.26&0.302&37&$-$18.6&21.1\\
10.2.23&10 44 04.00&+00 01 53.1&22.36&1.29&0.665&75&$-$20.1&22.9,23.8,24.0\\
13.2.10&13 42 04.80&+00 11 09.9&21.71& ---&0.424&~0&$-$20.8&21.1\\
13.2.13&13 42 05.53&+00 10 26.5&21.82&1.92&0.430&23&$-$20.0&21.7,24.0\\
13.2.22&13 42 06.40&+00 12 04.5&22.19&1.25&0.422&29&$-$19.0&21.8\\
13.4.12&13 41 13.65&+00 02 21.5&21.50& ---&0.120&38&$-$16.8&20.8\\
13.4.16&13 41 13.68&+00 00 31.2&21.72&0.95&0.120&56&$-$16.3&21.3\\
13.4.22&13 41 15.73&+00 00 51.9&22.10&0.67&0.086&36&$-$15.1&21.8\\
13.5.06&13 41 12.78&+00 08 06.0&22.40&1.25&0.112&54&$-$15.5&22.0,24.4\\
13.5.07&13 41 13.60&+00 08 29.5&21.32&0.56&0.220&24&$-$17.7&21.3\\
13.5.10&13 41 06.50&+00 12 19.8&21.57&1.45&0.329&31&$-$19.1&20.8\\
13.5.12&13 41 13.13&+00 11 45.5&21.63&1.19&0.678&27&$-$20.8&21.5\\
13.5.14&13 41 08.95&+00 12 52.5&21.83& ---&0.255&29&$-$18.8&21.1\\
13.5.20&13 41 06.72&+00 09 01.8&22.30&1.75&0.521&25&$-$19.9&21.6\\[10pt]
\multicolumn{9}{l}{(b) 1993 April 20--22} \\
ID&\multicolumn{2}{c}{R.A.~~~~(1950)~~~~Dec.}&b$_J$&b$_J$$-$r$_F$&
z&\EW(\AA)&~M$_{b_J}$&B (A,B,C)\\[10pt]
10.2.02&10 43 48.77&+00 07 38.6&21.40&0.73&0.549&~8&$-$19.4&21.2\\
10.2.04&10 43 55.81&+00 08 50.6&21.71&0.58&0.543&~0&$-$19.1&21.8\\
10.2.11&10 44 01.25&+00 09 41.5&21.03&1.24&0.277&27&$-$19.1&21.6,22.6\\
10.2.15&10 43 49.40&+00 08 11.9&21.70&1.77&0.451&~8&$-$20.1&22.3\\
10.2.20&10 43 45.16&+00 07 16.6&22.02&1.79&0.188&~0&$-$17.5&22.1\\
10.2.22&10 43 49.11&+00 11 09.3&22.11&0.89&0.180&31&$-$16.8&22.0\\
10.2.25&10 43 48.10&+00 06 38.0&22.43&0.90&0.160&20&$-$16.2&22.6\\
13.2.17&13 41 52.15&+00 03 07.1&22.08&1.42&0.202&~0&$-$17.4&22.5\\
13.2.23&13 41 59.64&+00 05 42.9&22.33&1.49&0.281&~0&$-$18.1&22.8\\
13.2.26&13 41 54.18&+00 03 26.0&22.41&0.64&0.598&~8&$-$18.6&22.5\\
13.2.36&13 41 58.30&+00 04 46.9&22.75&1.54&0.537&32&$-$19.4&22.9\\
\end{tabular}
\end{table*}

The large numbers and low redshifts of galaxies at faint apparent
magnitudes presents a serious problem for models of galaxy evolution.
Conventional models involving luminosity-independent luminosity
evolution can only match the number counts with a high redshift of
galaxy formation (z$_f$$>$5) and a low-density Universe
($\Omega$$\la$0.1). Even so they appear to be in conflict with the
lack of objects found at redshifts z$>$0.5, and in particular the low
redshifts found for the bluest objects with near-flat spectra
(f$_\nu$$\sim$constant), which such models predict ought to have z$>$1
(Colless \etal\ 1993). Better matches to the observed number counts
and redshift distributions are obtained by models which effectively
produce density evolution of the galaxy luminosity function.  Two
phenomenological models that give this result invoke, respectively, a
new population of star-forming dwarf galaxies that dominated the
galaxy population at moderate redshifts but which have since faded
beyond detection (Cowie \etal\ 1991, Babul \& Rees 1992), or strong
merging-driven evolution with a higher past star-formation rate which
may be directly associated with the dynamical interactions
(Rocca-Volmerange \& Guiderdoni 1990, Broadhurst \etal\ 1992).

A further feature of the evolution that needs to be accounted for in
any model is the increasing fraction of galaxies with large rest-frame
[OII] 3727\AA\ equivalent widths (\EW), indicative of enhanced
star-formation.  Measurements of \EW\ derived from the redshift
surveys of Broadhurst \etal\ (1988) and Colless \etal\ (1990, 1993),
together with a more recent survey at brighter magnitudes, shows
clearly that the steep slope of the galaxy counts is closely
associated with the increase in the number of galaxies undergoing
enhanced star-formation (see figure~2 of Broadhurst \etal\ 1992).
Moreover, the co-added spectra of the high-\EW\ galaxies suggests that
the star-formation is typically strong but of relatively short
($\sim$0.1~Gyr) duration (Broadhurst
\etal\ 1988).

What is the nature of the enhanced star-formation activity at moderate
redshifts? What are the physical mechanisms that produce it? Is it
triggered by galaxy interactions and/or mergers? Is it related to the
formation of disks or confined to the nucleus? The current
observational data (essentially counts, colours and redshifts) are not
sufficient to answer these questions, since the data could, in
principle, be reproduced by any process which leads to density
evolution of the luminosity function.  Qualitatively new information
is required to determine the cause and origin of the observed
evolution and decide which of the various physical processes that have
been proposed actually contribute to the increased star-formation
activity at moderate redshifts.

A promising approach is to take advantage of the proximity of those
star-forming galaxies that represent the count excess and use the best
ground-based images with seeing FWHM$\sim$0.5~arcsec to resolve
details on scales of $\la$2\kpc\ (H$_0$=100h~km~s$^{-1}$~Mpc$^{-1}$).
Sufficiently deep high-resolution images can therefore provide
independent measures of surface brightness at many points across the
face of moderate-redshift galaxies. By imaging in two colours the
light distributions for the star-forming (blue) component and the
quiescent stellar distribution (as revealed in the near infrared) can
be compared to establish whether the star formation is localised in
the nucleus or disk or uniformly distributed across the entire galaxy,
and whether it is triggered by tidal interactions (signalled by
distorted isophotes and/or close companions).

An initial step in this direction was taken by Giraud (1992) who
observed 30 faint (22$<$V$<$24.5), flat-spectrum (V$-$I$\leq$0.8), low
surface brightness ($\overline{\mu_V}$$>$23) galaxies. Using the NTT he
obtained V and I images with seeing FWHM better than 0.85~arcsec in V
and 0.7~arcsec in I. The galaxies (whose redshifts are not known) were
found to fall into three broad classes: compact (having a dominant point
source and fainter envelope), irregular (elongated, spiral or distorted)
and multiple (more than one peak in surface brightness). The relative
numbers of objects in these classes were approximately 2:3:1.

This paper reports the results of a programme of high-resolution
mulitcolour imaging of star-forming galaxies {\em with measured
redshifts and [OII] equivalent widths} from the survey of Colless
\etal\ (1990, 1993). Section~2 describes the selection of the target
galaxies and the observations. Section~3 outlines the method used to
obtain scalelengths for the galaxies and gives a detailed description
of each object. The galaxies' morphologies, their size--luminosity
relation and the incidence of close companions are discussed in
Section~4 in relation to evolutionary models. Our conclusions are
presented in Section~5.

\section{OBSERVATIONS}

The galaxies selected for imaging were taken from the deep redshift
survey of Colless \etal\ (1990, 1993). They fall into two classes: a
primary sample of 17 galaxies with [OII] 3727\AA\ equivalent widths
\EW$>$20\AA, indicative of enhanced star-formation, and a comparison
sample of 9 galaxies with \EW$<$10\AA. Details of the galaxies
observed are given in Table~1, which lists identification number from
the redshift survey, R.A.\ and Dec.\ (1950), b$_J$ magnitude,
b$_J$$-$r$_F$ colour, redshift, [OII] 3727\AA\ \EW, and approximate
absolute b$_J$ magnitude.  Fig.~1 displays the distributions of
apparent magnitude, absolute magnitude and \EW\ for the galaxies
imaged in this study, and shows that they are statistically
representative of the galaxies in the redshift survey.

\begin{figure}  % sfa1_fig1.sm & .ps
\centering
\vspace{11cm}
\caption{The distributions of apparent magnitude, absolute magnitude
and [OII] \EW\ for all the galaxies identified in the Colless
\etal\ redshift survey sample (open histogram), and for the galaxies
selected for imaging (hatched histogram).}
\end{figure}

The observations were made with HRCam on the Canada-France-Hawaii
Telescope (CFHT) during the nights of 1992 April 7--8 and 1993 April
20--22. HRCam is a high-resolution camera which can fast-guide at
10--200~Hz using a bright nearby guide star in order to achieve
improved image quality over a 2.2~arcmin diameter field. A detailed
description is given by McClure \etal\ (1989).  During the first run
we used the SAIC1 1024$^2$ CCD, which gave a pixel size of
0.13~arcsec.  The I filter had $\lambda_c$=8310\AA\ and
$\Delta\lambda$=1970\AA; the V filter used had $\lambda_c$=5420\AA\
and $\Delta\lambda$=900\AA (we were obliged to use V rather than B due
to this CCD's poor blue response). For the second run we were able to
image in B and I using the Loral~3 2048$^2$ CCD, which has 0.11~arcsec
pixels. The B filter used in this run had $\lambda_c$=4300\AA\ and
$\Delta\lambda$=970\AA, while the I filter was the same as for the
previous run.

All the observations are summarised in Table~2, which gives, for each
field, the targets in that field and the total integration time and
seeing FWHM for the images in each filter. Where there were two series
of exposures with slightly different pointings or seeing they were
co-added separately. The seeing point spread functions (PSFs) were
measured from a few stars on each image. All fields from the first run
have V and I images with effective seeing FWHM of 0.5--0.7~arcsec,
with the exception of 13F5 which has FWHM of 1.0~arcsec in V and
1.2~arcsec in I. The fields from the second run have B and I images
with FWHM of 0.6--1.0~arcsec, with the exception of the B image of
13F7, which has a FWHM of 1.5~arcsec.

\begin{table}
\centering
\caption{Log of observations.}
\begin{tabular}{cccccc}
\multicolumn{6}{l}{(a) 1992 April 7--8} \\
Field & Objects  & I           & V1          & V2    & \\[10pt]
10F1  & 10.2.05  & 1800s       & 2700s       & 2700s       & \\
      & 10.2.17  & 0.56\arcsec & 0.69\arcsec & 0.72\arcsec & \\
      & 10.2.23  &             &             &             & \\
13F1  & 13.4.12  & 1800s       & 3000s       & 1800s       & \\
      & 13.4.16  & 0.59\arcsec & 0.91\arcsec & 0.55\arcsec & \\
      & 13.4.22  &             &             &             & \\
13F2  & 13.5.06  & 1800s       & 1300s       & 3600s       & \\
      & 13.5.07  & 0.59\arcsec & 0.55\arcsec & 0.57\arcsec & \\
      & 13.5.20  &             &             &             & \\
13F3  & 13.5.10  & 1800s       & 5400s       &             & \\
      & 13.5.12  & 0.62\arcsec & 0.70\arcsec &             & \\
      & 13.5.14  &             &             &             & \\
13F5  & 13.2.10  & 1800s       & 5400s       &             & \\
      & 13.2.13  & 1.21\arcsec & 1.01\arcsec &             & \\
      & 13.2.22  &             &             &             & \\[10pt]
\multicolumn{6}{l}{(b) 1993 April 20--22} \\
Field & Objects  & B1          & B2          & I1          & I2    \\[10pt]
10F2  & 10.2.02  & 5000s       & 5000s       & 2000s       & 2000s       \\
      & 10.2.15  & 1.00\arcsec & 0.80\arcsec & 0.90\arcsec & 0.60\arcsec \\
      & 10.2.20  &             &             &             &             \\
      & 10.2.25  &             &             &             &             \\
10F3  & 10.2.04  & 5000s       &             & 2000s       &             \\
      & 10.2.11  & 1.00\arcsec &             & 0.70\arcsec &             \\
      & 10.2.22  &             &             &             &             \\
13F4  & 13.2.17  & 5000s       &             & 2000s       &             \\
      & 13.2.26  & 0.80\arcsec &             & 0.80\arcsec &             \\
13F7  & 13.2.23  & 5000\arcsec &             & 2000s       &             \\
      & 13.2.36  & 1.50\arcsec &             & 1.00\arcsec &             \\
\end{tabular}
\end{table}

Although these images were not taken in photometric conditions we have
approximately zero-pointed the B and V magnitudes by one-to-one
comparison of the objects on each image with the b$_J$ and r$_F$
photometry from Colless \etal\ (based on Jones \etal\ 1991).  The B and
V magnitudes thus derived are given in Table~1 for each of the resolved
objects close to the position of the target galaxy. These magnitudes
have an rms error of 0.2~mag.

Fig.~2 shows the V and I or B and I images in a small region about each
target galaxy. The seeing FWHM is shown as the small vertical bar on
each image (see also Table~2), while the larger horizontal bar on the
leftmost image corresponds to 5\kpc\ at the redshift of the galaxy. In
the cases where there is more than one resolved object close to the
position of the target galaxy, these are labelled A, B, C in decreasing
order of brightness.

\begin{figure*}  % sfa1_fig2a.ps
\centering
\vspace{22cm}
\caption{(a) V and I images in a region 7.9~arcsec (61~pixels) square
about each target galaxy from the 1992 observations. The I image is at
left, with the V1 and V2 images to the right. The seeing FWHM is shown as
the small vertical bar on each image; the horizontal bar on the I image
corresponds to 5\kpc\ at the redshift of the galaxy.}
\end{figure*}

\begin{figure*}  % sfa1_fig2b.ps
\centering
\vspace{22cm}
\caption*{(b) B and I images in a region 6.7~arcsec (61~pixels) square
about each target galaxy from the 1993 observations. The B1 and B2
images are at left, with the I1 and I2 images to the right. The seeing
FWHM is shown as the small vertical bar on each image; the horizontal
bar on the B1 image corresponds to 5\kpc\ at the redshift of the
galaxy.}
\end{figure*}

Five of the objects in the sample (10.2.11, 10.2.23, 13.2.13, 13.5.06
and 13.5.07) are seen in Fig.~2 to have close companions. The closer
pairs were not resolved in the b$_J$ and r$_F$ photometry (based on
prime focus AAT plates taken in 1.5--2.5 arcsec seeing), so that the
b$_J$ and r$_F$ magnitudes for these objects in Table~1 refer to the
pair combined. The B and V magnitudes in Table~1, however, are for each
object individually, and in general there is a magnitude or more
difference between objects A and B. We may therefore reasonably
attribute the spectroscopically-measured parameters (redshifts and
equivalent widths) to the brighter object. The exception is 13.5.07,
where the objects are still only partially resolved in 0.6~arcsec
seeing, and also appear more similar in magnitude; only a combined V
magnitude is given in Table~1.

\section{ANALYSIS}

The resolution of the images in Fig.~2 is insufficient to permit a
direct visual classification of the galaxies' morphologies. We can
still obtain useful information, however, from the overall form and
extent of their light profiles. Although we cannot achieve a full
bulge/disk decomposition, we can say whether individual objects are
best fit by exponential disks, $r^{1/4}$-law bulges or as point
sources. We can also establish a scalelength for each galaxy and obtain
a size-luminosity relation. Finally, by comparing the scalelengths
obtained in our two colours, we may obtain an indication of the
relative importance of the bulge and disk in each object.

In order to carry out this analysis, we have fitted two-dimensional
exponential disks and $r^{1/4}$ bulges directly to the images of Fig.~2
using $\chi^2$-minimisation. Because seeing effects are significant,
before fitting we have convolved the models with a stellar profile
either taken from the same frame or from a frame with very similar
seeing FWHM. In the case of close pairs of objects only the brighter
was fitted, with the companion's image excised. Fig.~3 compares the
surface brightness profiles (in circular apertures) of the galaxies and
the best-fitting models. The B and V surface brightness scales in the
figure are based on the calibration described in Section~2; the I
calibration is only approximate. The models generally fit the profiles
within the errors down to a surface brightness in B and V of
$\mu\approx25$--26~mag~arcsec$^{-2}\approx29.5$--30.5~mag~pixel$^{-2}$.
These limits typically correspond to between three and five times the
galaxy's scalelength. The results of the fits are presented in Table~3,
which lists the parameters of the best-fitting models: the disk
scalelength in both arcsec and \kpc\ (or effective radius where a bulge
fitted significantly better, indicated in the table by an asterisk),
and the axial ratio.

\begin{figure*}  % sfa1_fig3a.sm & .ps
\centering
\vspace{22cm}
\caption{(a) Surface brightness profiles in circular apertures for each
target galaxy from the 1992 observations (points with error bars),
together with the best-fitting model including seeing convolution
(dotted lines).}
\end{figure*}

\begin{figure*}
\centering
\vspace{22cm}
\caption*{(a) -- {\it continued}}
\end{figure*}

\begin{figure*}  % sfa1_fig3b.sm & .ps
\centering
\vspace{22cm}
\caption*{(b) Surface brightness profiles in circular apertures for
each target galaxy from the 1993 observations (points with error bars),
together with the best-fitting model including seeing convolution
(dotted lines). Note that there is no plot for 10.2.11 I1 because the
galaxy was too near the edge of the frame (see Fig~2b).}
\end{figure*}

\begin{figure*}
\centering
\vspace{22cm}
\caption*{(b) -- {\it continued}}
\end{figure*}

\begin{table*}
\centering
\caption{Disk scalelengths and axial ratios.}
\begin{tabular}{rccclllrccc}
   Galaxy & \multicolumn{2}{c}{Scalelength} & Axial       &&&& Galaxy     &
\multicolumn{2}{c}{Scalelength} & Axial       \\
    image & (arcsec)      & (h$^{-1}$~kpc)  & ratio       &&&&  image     &
(arcsec)      & (h$^{-1}$~kpc)  & ratio       \\[10pt]
10.2.05 I & $<$0.05       &     ---         &     ---     &&&& 10.2.02 I1 &
$<$0.05       &     ---         &     ---     \\
       V1 & $<$0.05       &     ---         &     ---     &&&&         I2 &
$<$0.05       &     ---         &     ---     \\
       V2 & $<$0.05       &     ---         &     ---     &&&&         B1 &
$<$0.05       &     ---         &     ---     \\
10.2.17 I & 0.37 (0.02)   & 1.02 (0.06)     & 0.68 (0.03) &&&&         B2 &
$<$0.05       &     ---         &     ---     \\
       V1 & 0.34 (0.01)   & 0.94 (0.03)     & 0.69 (0.04) &&&& 10.2.04 I1 &
$<$0.05       &     ---         &     ---     \\
       V2 & 0.38 (0.01)   & 1.05 (0.03)     & 0.73 (0.04) &&&&         B1 &
$<$0.2        &   $<$0.7        &     ---     \\
10.2.23 I & $<$0.2        &   $<$0.8        &     ---     &&&& 10.2.11 I1 & ---
          &     ---         &     ---     \\
       V1 & $<$0.2        &   $<$0.8        &     ---     &&&&         B1 &
0.74 (0.06)   & 1.94 (0.16)     & 0.90 (0.10) \\
13.2.10 I & 0.45 (0.01)   & 1.49 (0.03)     & 0.74 (0.03) &&&& 10.2.15 I1 &
1.28 (0.11)*  & 4.35 (0.37)*    & 0.63 (0.06) \\
       V1 & 0.40 (0.02)   & 1.32 (0.07)     & 0.64 (0.04) &&&&         I2 &
1.05 (0.04)*  & 3.57 (0.14)*    & 0.75 (0.03) \\
13.2.13 I & 0.85 (0.12)   & 2.83 (0.40)     & 0.35 (0.06) &&&&         B1 &
0.53 (0.09)*  & 1.80 (0.31)*    & 0.82 (0.16) \\
       V1 & 0.36 (0.02)   & 1.20 (0.07)     & 0.37 (0.06) &&&&         B2 &
0.61 (0.04)*  & 2.08 (0.14)*    & 0.65 (0.10) \\
13.2.22 I & 0.38 (0.08)   & 1.26 (0.26)     & 0.66 (0.24) &&&& 10.2.20 I1 &
1.12 (0.32)*  & 2.26 (0.65)*    & 0.56 (0.06) \\
       V1 & 0.59 (0.05)   & 1.95 (0.17)     & 0.70 (0.07) &&&&         I2 &
0.96 (0.06)*  & 1.94 (0.12)*    & 0.46 (0.05) \\
13.4.12 I & 0.19 (0.06)   & 0.27 (0.09)     & 0.09 (0.03) &&&&         B1 &
0.74 (0.18)*  & 1.50 (0.36)*    & 0.34 (0.07) \\
       V1 & 0.22 (0.02)   & 0.31 (0.03)     & 0.02 (0.01) &&&&         B2 &
0.37 (0.09)*  & 0.75 (0.18)*    & 0.77 (0.10) \\
       V2 & 0.22 (0.01)   & 0.31 (0.02)     & 0.03 (0.02) &&&& 10.2.22 I1 &
0.47 (0.02)   & 0.92 (0.04)     & 0.90 (0.06) \\
13.4.16 I & 0.59 (0.03)   & 0.84 (0.04)     & 0.47 (0.03) &&&&         B1 &
0.50 (0.01)   & 0.98 (0.02)     & 0.90 (0.30) \\
       V1 & 0.67 (0.04)   & 0.96 (0.06)     & 0.54 (0.05) &&&& 10.2.25 I1 &
0.74 (0.10)   & 1.32 (0.18)     & 0.78 (0.14) \\
       V2 & 0.71 (0.04)   & 1.01 (0.06)     & 0.51 (0.04) &&&&         I2 &
0.57 (0.12)   & 1.02 (0.22)     & 0.70 (0.10) \\
13.4.22 I & 0.50 (0.03)   & 0.54 (0.03)     & 0.42 (0.05) &&&&         B1 &
0.55 (0.11)   & 0.98 (0.20)     & 0.76 (0.17) \\
       V1 & 0.48 (0.04)   & 0.52 (0.04)     & 0.44 (0.07) &&&&         B2 &
0.63 (0.04)   & 1.08 (0.07)     & 0.60 (0.10) \\
       V2 & 0.48 (0.03)   & 0.52 (0.03)     & 0.29 (0.04) &&&& 13.2.17 I1 &
$<$0.05       &     ---         &     ---     \\
13.5.06 I & $<$0.05       &     ---         &     ---     &&&&         B1 &
$<$0.1        &     ---         &     ---     \\
       V1 & $<$0.05       &     ---         &     ---     &&&& 13.2.23 I1 &
0.41 (0.04)   & 1.08 (0.11)     & 0.52 (0.10) \\
       V2 & $<$0.05       &     ---         &     ---     &&&&         B1 &
0.47 (0.11)   & 1.24 (0.29)     & 0.72 (0.20) \\
13.5.07 I & 0.32 (0.02)   & 0.72 (0.05)     & 0.26 (0.06) &&&& 13.2.26 I1 &
$<$0.1        &     ---         &     ---     \\
       V1 & 0.35 (0.02)   & 0.79 (0.05)     & 0.19 (0.05) &&&&         B1 &
$<$0.1        &     ---         &     ---     \\
       V2 & 0.35 (0.01)   & 0.79 (0.03)     & 0.17 (0.03) &&&& 13.2.36 I1 &
0.17 (0.07)   & 0.62 (0.26)     & 0.08 (1.00) \\
13.5.10 I & 0.71 (0.03)   & 2.06 (0.09)     & 0.71 (0.04) &&&&         B1 &
0.50 (0.12)   & 1.83 (0.44)     & 0.57 (0.2)  \\
       V1 & 0.92 (0.03)   & 2.67 (0.09)     & 0.64 (0.02) &&&&
          &                 &             \\
13.5.12 I & 0.77 (0.06)   & 3.03 (0.24)     & 0.66 (0.07) &&&&
          &                 &             \\
       V1 & 1.54 (0.13)   & 6.06 (0.51)     & 0.52 (0.04) &&&&
          &                 &             \\
13.5.14 I & 0.53 (0.04)   & 1.32 (0.10)     & 0.74 (0.08) &&&&
          &                 &             \\
       V1 & 0.83 (0.03)   & 2.06 (0.07)     & 0.73 (0.03) &&&&
          &                 &             \\
13.5.20 I & 0.45 (0.03)   & 1.63 (0.11)     & 0.99 (0.09) &&&&
          &                 &             \\
       V1 & 0.59 (0.06)   & 2.14 (0.22)     & 0.90 (0.12) &&&&
          &                 &             \\
       V2 & 0.62 (0.04)   & 2.25 (0.14)     & 0.85 (0.08) &&&&
          &                 &             \\
\end{tabular}
\end{table*}

In order to check the results of the fitting procedure we simulated
galaxy images bracketing the range of S/N and structural parameters
observed in our data. We constructed sets of model disk galaxies with
scalelengths ranging from 0.05~arcsec for the smallest set up to
1.0~arcsec for the largest, and sets of model bulges with effective
radii from 0.45~arcsec to 2.5~arcsec. Each set consisted of 36 models.
The total counts in the model galaxies were matched to those of the
observed galaxies. The models were placed at random locations with
respect to pixel centres, and we convolved them with stellar profiles
from our data which spanned the range in seeing of our observations.
Finally we added sky and Poisson noise.

These simulated images were then analysed in the same way as the data.
We measured the galaxy centroids and the sky levels from the images
(rather than fixing them at their known values), then applied our
fitting procedure. The fitted parameters for each set of 36 models were
then compared to the input parameters to estimate the distribution of
errors. We found that these distributions were approximately normal,
and that the errors given by the fitting routine were reliable in that
on average approximately 68\% of the true values were within the
estimated error of the fitted value and approximately 95\% were within
twice the estimated error. We also found that, over the range of seeing
in our observations, the errors in the scale sizes increased only
slowly with increasing FWHM.

These simulations, which mimic as closely as possible the properties of
the galaxies and noise in our data, thus give us confidence in the
errors we estimate for the fitted parameters. As Table~3 shows, these
errors are generally small compared to the galaxies' scalelengths. We
conclude that the depth and spatial resolution of our images are
sufficient to derive reliable scalelengths for all but the smallest of
our target galaxies.

As well as fitting simulated disk galaxies with disk models and
simulated bulge galaxies with bulge models, we also fitted disks with
bulges and bulges with disks, since for real galaxy images we have no
way of knowing in advance the appropriate model to fit. We found that,
for the range of scale sizes, S/N and seeing FWHM in our observations,
the discrimination between a disk galaxy and a bulge galaxy was
considerably better in those cases where the ratio of FWHM to scale
size was smaller. This improved ability to discriminate disks and
bulges was the main benefit derived from the reduction in the seeing
FWHM produced by HRCam's fast guiding. Although our data are {\em not}
adequate for quantitative measurement of bulge-to-disk ratios, in most
cases they {\em are} adequate to discriminate between bulge-dominated
galaxies and disk-dominated galaxies.

In fact most of the galaxies in the sample can be adequately fit by an
exponential disk. Only two galaxies (10.2.15 and 10.2.20) were
significantly better fit by an r$^{1/4}$ bulge, while seven objects
could not be satisfactorily fitted because they were either unresolved
or only  marginally resolved. In the latter case the objects could be
equally well fitted by a disk or a bulge because the light
distribution, though extended, was still dominated by the seeing PSF.
Only upper limits to the scalelengths of these seven objects are given
in Table~3, and these limits are also uncertain. They include 5 of the
9 low-\EW\ objects and just 2 of the 17 high-\EW\ objects (10.2.23 and
13.5.06), both of which have close companions and very large \EW. There
is excellent agreement between the fitted parameters obtained from
repeat images in the same passband, and a correlation between the V and
I or B and I scalelengths, with the blue scalelengths tending to be
larger than the red.

In Fig.~4 the scalelengths of the sample galaxies, as measured
from both the blue and red images, are plotted against their absolute
b$_J$ magnitudes. For comparison, the same diagram is plotted for a
sample of low-redshift galaxies with V and I scalelengths measured by
Ryder (1993) from CCD surface photometry and absolute B magnitudes from
the Nearby Galaxies Catalogue (Tully 1988; converted to h=1 and
transformed to b$_J$ using b$_J$$\approx$B$-$0.25). The nearby and
distant galaxies both populate the same relatively narrow locus on this
diagram, which is consistent with Holmberg's (1975) relation that
M$\propto$$-$6.0log(size), indicated on the figure by the dotted line.

\begin{figure}  % sfa1_fig4.sm\.ps
\centering
\vspace{11cm}
\caption{Size versus luminosity, showing the log of the galaxies' disk
scalelengths (in kpc), measured from the B or V images and the I
images, against their absolute b$_J$ magnitudes (assuming
H$_0$=100~km~s$^{-1}$~Mpc$^{-1}$). Estimated errors on the fitted
scalelengths are shown. For comparison the same relation is plotted
(small squares) for a sample of nearby galaxies (Ryder 1993). The
dotted line corresponds to a Holmberg-type luminosity--size relation
(Holmberg 1975) in which M$\propto$$-$6.0log(size).}
\end{figure}

\section{DESCRIPTIONS OF INDIVIDUAL OBJECTS}

Combining the redshifts, equivalent widths, colours and absolute
magnitudes from the Colless \etal\ redshift survey with the new imaging
data, we describe in this section the morphological type and mode of
star formation for each of the 26 galaxies in the sample. The main
results and trends drawn from this detailed analysis are discussed in
the following section.

{\em 10.2.05:} This M$_{b_J}$=$-$19.2 object has a low [OII] equivalent
width (\EW=4\AA) and a colour of b$_J$$-$r$_F$=1.91, consistent with an
elliptical at its redshift of z=0.303 (see, \eg, figure~12 of Colless
\etal\ 1990). However it appears to be unusually compact, since an
elliptical galaxy of this absolute magnitude would typically have an
effective radius of $\sim$2.5\kpc\ (see, \eg, figure~4 of Sandage \&
Perelmutter 1990) and should appear extended, whereas this object is
unresolved. It unresolved nature and colour might also be consistent
with a late-type star, however its spectrum is unambiguously that of an
early-type galaxy at z=0.303.

{\em 10.2.17:} At the same redshift as 10.2.05, this M$_{b_J}$=$-$18.6
object is well-fit by an exponential disk with a scalelength of
1\kpc\ (in both V and I). It has b$_J$$-$r$_F$=1.26 and \EW=37\AA, and
is generally consistent with a mid-type spiral galaxy with enhanced
star-formation across its entire disk.

{\em 10.2.23:} This is the second most distant galaxy in the sample, at
z=0.665, and has the highest [OII] equivalent width (\EW=75\AA). It has
a redder companion (B in Fig.~2) at a projected separation of 5.3\kpc.
The pair were not resolved in the original photographic image, so that
the main object (A) is both fainter and bluer than that photometry
would indicate. Since the V magnitudes of A and B differ by 0.9~mag
(Table~1), the estimated absolute magnitude for A of M$_{b_J}$=$-$20.1
is probably too bright by $\sim$0.4~mag. The light distribution
of A is very compact, having a scalelength of less than 0.2~arcsec
($<$0.8\kpc) which is small for its luminosity. Subtracting a point
source from this object shows no convincing residuals in I, but in V
there is a significant residual consistent with an underlying disk.
This object thus appears to be a sub-L$^\ast$ spiral whose light is
dominated by a nuclear starburst, possibly triggered by tidal
interaction with the close companion.

{\em 13.2.10:} This object at z=0.424 has no detected [OII] emission
and is poorly fit by a pure exponential disk due to a significant bulge
component (see Fig.~3). It is one of the intrinsically brightest
objects in the sample at M$_{b_j}$=$-$20.8 (\ie\ $\sim$1~mag brighter
than M$^\ast$) and is probably a giant early-type spiral or S0.

{\em 13.2.13:} The main object (A in Fig.~2) has a much redder
companion (B) at a projected separation of 6.3\kpc. As with 10.2.23,
the two objects were not resolved in the photographic photometry,
however B, though prominent in I, is 2.3~mag fainter than A in V.  The
measured absolute magnitude of M$_{b_j}$=$-$20.0 is thus not likely to
be much altered, although the red b$_J$$-$r$_F$ colour is due to the
very red companion. The disk scalelength fitted in I is over twice that
found in V, probably because of inadequate removal of contaminating
light from the companion. The V estimate of 1.2\kpc\ is therefore
preferred. The galaxy appears to be an M$_\ast$ spiral with moderate
star-formation (\EW=23\AA) over its whole disk. It is possible the
star-formation is related to the presence of a close companion, but if
so the burst is mild and global, similar to 10.2.11 and unlike 10.2.23
or 13.5.06 where the evidence links the close companions to strong
star-formation within a small region, probably just the nucleus.

{\em 13.2.22:} This object, at a similar redshift (z=0.422) to both
13.2.10 and 13.2.13, has \EW=29\AA. The V scalelength is 50\% greater
than that in I, implying that the old stellar population in this
sub-L$^\ast$ galaxy is more concentrated than the star-forming regions.
Together with a colour of b$_J$$-$r$_F$=1.25, this suggests the galaxy
is a mid- to late-type spiral with moderate star-formation in the
disk.

{\em 13.4.12:} Like 10.2.23, this object, though clearly extended, does
not yield very reliable fits, mainly because it is very flat (axial
ratio $\sim$0.1). At z=0.120 it is one of the closer objects in the
sample, so that the scalelength of 0.2~arcsec corresponds to just
0.3\kpc. Even though it is a relatively faint galaxy
(M$_{b_j}$=$-$16.8), this scalelength puts it well below the mean
relation in Fig.~4. The data are consistent with an edge-on spiral
galaxy undergoing a strong burst of star-formation (\EW=38\AA). This
star-formation is unlikely to be concentrated in the nucleus, as
the object would then appear round rather than flat.

{\em 13.4.16:} This object provides an interesting contrast to
13.4.12.  Although at the same redshift and with a similarly high [OII]
equivalent width (\EW=56\AA) and faint absolute magnitude
(M$_{b_J}$=$-$16.3), it has a V disk scalelength of 1\kpc, larger than
might be expected for such a faint galaxy (see Fig.~4). Comparison of
the V and I images shows that the blue and red stellar populations are
co-extensive and that the strong star-formation indicated by \EW\ and
the colour of b$_J$$-$r$_F$=0.95 is occurring over the whole disk of
what is probably a late-type spiral.

{\em 13.4.22:} At z=0.086 this is the lowest redshift object in the
sample and has the faintest absolute magnitude (M$_{b_j}$=$-$15.1). It
also has the most irregular isophotes, although both the V and I images
are well-fit by an exponential disk of scalelength 0.5\kpc. This puts
it slightly above the extrapolated ridgeline of the size-luminosity
relation (see Fig.~4). It has \EW=36\AA\ and b$_J$$-$r$_F$=0.67, and is
probably a star-forming irregular galaxy.

{\em 13.5.06:} This object is similar to 13.4.22, having z=0.112,
M$_{b_j}$=$-$15.5 and \EW=54\AA. However it has a much redder companion
at a projected separation of just 1.9\kpc\ which was not resolved in
the photographic image, explaining the colour of b$_J$$-$r$_F$=1.25.
The main object (A in Fig.~2) is unresolved even in the HRCam images,
so that it is a good candidate for a strong burst of (possibly nuclear)
star-formation in a dwarf galaxy, which may have been triggered by the
close companion.

{\em 13.5.07:} This object is in fact a very close  pair of galaxies
(separation 1.7\kpc) that are only just resolved in the HRCam images.
The pair overlap too much to permit separate photometry of the
components, and in fact it is possible to fit an exponential disk
(scalelength 0.8\kpc) to the combined system, although the fits to some
of the images are poor and the axial ratio is just 0.2. At z=0.220, the
pair have a combined luminosity of M$_{b_J}$=$-$17.7. The colour of the
combined system is very blue (b$_J$$-$r$_F$=0.56) but the [OII]
emission (\EW=24\AA) is relatively weak. If tidal interactions are
causing the star formation, then this system more closely resembles
13.2.13 than the probable nuclear starbursts in 10.2.23 and 13.5.06.

{\em 13.5.10:} With z=0.329 and M$_{b_J}$=$-$19.1, this galaxy is
similar to 10.2.17. It has a red colour (b$_J$$-$r$_F$=1.45) for its
[OII] equivalent width of 31\AA, and is not well fit by an exponential
disk. The data are consistent with an early-type spiral having a
significant bulge component and star-formation in its disk.

{\em 13.5.12:} At z=0.678 this is the most distant object in the
sample. It is also the biggest and (with 13.2.10) the brightest, having
M$_{b_J}$=$-$20.8. The V image is well-fit by a disk with a scalelength
of 6\kpc, however the I image is not so well fit, due to a more
significant bulge, and has a scalelength of just 3\kpc.  The colour of
b$_J$$-$r$_F$=1.19 and equivalent width of \EW=27\AA\ suggest moderate
star-formation is occurring over a very extended disk in an unusually
bright mid-type spiral.

{\em 13.5.14:} As with 13.5.12, this galaxy has a larger V than I
scalelength (2\kpc compared to 1.3\kpc) and is not as well fitted by a
disk in I. It has M$_{b_J}$=$-$18.8 and, with an [OII] equivalent width
of 29\AA, appears to be a mid-type spiral with moderate star formation
occurring across its disk.

{\em 13.5.20:} This object at z=0.521 has quite a red colour
(b$_J$$-$r$_F$=1.75) but an equivalent width of 25\AA. It is well-fit
in both V and I by a pure disk with only a 30\% greater scalelength in
V than I. It has a luminosity close to L$^\ast$, and is probably an
early-type spiral with star-formation in its disk.

{\em 10.2.02:} This object has a redshift of z=0.549 and an absolute
magnitude of M$_{b_J}$=$-$19.4. It is very compact in both the B and I
images, being essentially unresolved. Given its very blue colour
(b$_J$$-$r$_F$=0.73), it has a low [OII] equivalent width (\EW=8\AA).

{\em 10.2.04:} Also unresolved in both B and I, this object has similar
redshift (z=0.543), absolute magnitude (M$_{b_J}$=$-$19.1) and colour
(b$_J$$-$r$_F$=0.58) to 10.2.02, although it has no detected [OII].

{\em 10.2.11:} This object at z=0.277 has a companion one magnitude
fainter in B at a projected distance of 5\kpc. The two objects are
clearly resolved although their isophotes overlap. The brighter (A in
Fig.~2) is adequately represented by a disk of scalelength 1.9\kpc\ in
B. Unfortunately the object fell at the very edge of the I frame and so
the red scalelength could not be obtained. The combined absolute
magnitude and colour of the pair are M$_{b_J}$=$-$19.1 and
b$_J$$-$r$_F$=1.24. The [OII] equivalent width of \EW=27\AA\ and the
smooth distribution of blue light across the disks of both objects
imply that if the star-formation is related to interaction between the
pair, it is relatively mild and global (similar to 13.2.13) rather than
a strong nuclear burst (as in 10.2.23 and 13.5.06).

{\em 10.2.15:} This object is marginally better fit by an $r^{1/4}$
bulge than an exponential disk in both B and I. Subtracting the
best-fit bulge leaves a faint indication of extended structure. The
image appears to be that of an M$^\ast$ bulge-dominated spiral, which
is consistent with the observed colour of b$_J$$-$r$_F$=1.77 (for the
object's redshift of z=0.451) and its low [OII] equivalent width
(\EW=8\AA).

{\em 10.2.20:} As with 10.2.15, this object is better fit by a bulge
than a disk in both B and I. There is faint residual structure that
might be an edge-on disk. Again, the colour (b$_J$$-$r$_F$=1.79) is
consistent with a bulge-dominated spiral at the object's redshift of
z=0.188. However the absolute magnitude (M$_{b_J}$=$-$17.5) is
considerably fainter than that of 10.2.15. No [OII] emission is
detected in this galaxy.

{\em 10.2.22:} At a redshift of z=0.180, this galaxy has
M$_{b_J}$=$-$16.8 and b$_J$$-$r$_F$=0.89. The blue colour accords with
its strong [OII] emission (\EW=31\AA). It is fit in both B and I by a
disk with scalelength 0.9\kpc, and is of roughly typical size for its
luminosity (Fig.~4). This object appears to be a strongly star-forming
late-type dwarf galaxy.

{\em 10.2.25:} This object is very similar to 10.2.22, though somewhat
fainter at M$_{b_J}$=$-$16.2. It has z=0.160 and b$_J$$-$r$_F$=0.90,
and an [OII] equivalent width of \EW=20\AA. It is slightly larger than
10.2.22, with a B and I scalelength of 1.0\kpc, making it somewhat
larger than usual given its fainter luminosity. It too appears to be a
late-type dwarf galaxy with enhanced star-formation.

{\em 13.2.17:} This object is effectively unresolved in both B and I.
It has a redshift of z=0.202 and an absolute magnitude of
M$_{b_J}$=$-$17.4. Its colour is  b$_J$$-$r$_F$=1.42 which corresponds
to a bulge-dominated spiral at this redshift. No [OII] emission is
detected.

{\em 13.2.23:} Although it has a similar redshift (z=0.281) and colour
(b$_J$$-$r$_F$=1.49) to 13.2.17, and also has no [OII], this object is
nearly a magnitude brighter, at M$_{b_J}$=$-$18.1. It is also well-fit
in both B and I by an exponential disk of scalelength 1.1--1.2\kpc,
exactly as expected for a galaxy of this luminosity, and is in all
respects consistent with being an early-type spiral.

{\em 13.2.26:} This high-redshift (z=0.598) galaxy is similar to
10.2.02, having M$_{b_J}$=$-$18.6, a low [OII] equivalent width
(\EW=8\AA) and a very blue colour (b$_J$$-$r$_F$=0.64). It is also
unresolved, although both the B and I images of this object were taken
in relatively poor (1--1.5~arcsec) seeing.

{\em 13.2.36:} This object has z=0.537, M$_{b_J}$=$-$19.4 and
b$_J$$-$r$_F$=1.54. It is well-fit in B by a disk of scalelength
1.8\kpc, consistent with its luminosity, but in I by a disk of
scalelength 0.6\kpc. The colour, large [OII] equivalent width
(\EW=32\AA), and the larger blue than red scalelength suggest that this
was a relatively small faint galaxy that has been brightened by strong
star-formation over the whole of an extensive disk. It is thus similar
to 13.5.12.

\section{DISCUSSION}

There is considerable morphological variation amongst the objects in
both the high-\EW\ and low-\EW\ samples. A visual impression of this
morphological variety can be obtained by comparing in Fig.~2 the
objects with similar redshifts, such as 13.4.12 and 13.4.16 (both
z=0.120), 10.2.05 and 10.2.17 (z=0.303,0.302), and 10.2.23 and 13.5.12
(z=0.665,0.678).

Comparison of the blue and red images does not in general show
significant variations in colour across the face of individual galaxies
on small scales (comparable to the seeing disk, \ie\ 1--2\kpc),
although to be detectable the variation in colour would have to be of
the order of 0.5~mag on these scales. Most (15/17) of the
high-\EW\ objects apparently have star-formation occurring globally,
indicated both by the relatively smooth distribution of the blue light
and by a blue scalelength comparable to or larger than the red
scalelength. Only two objects (10.2.23 and 13.5.06) show evidence for
nuclear starbursts. The former is the second most distant object in the
sample, at z=0.665, while the latter is one of the closer objects, at
z=0.112. Both objects have very large [OII] equivalent widths and
close companions, and are particularly strong cases for
interaction-triggered star-formation. Other objects with similarly
large \EW\ (\eg\ 10.2.17 and 13.4.16) do {\em not} have close
companions, while the remaining three galaxies {\em with} close
companions (13.2.13, 13.5.07 and 10.2.11) have \EW=20--30\AA and show
apparently global star-formation.

The scalelengths of the extended objects vary from 0.3\kpc\ for the
compact 13.4.12 (\EW=38\AA) at z=0.120, to 6\kpc\ for 13.5.12
(\EW=27\AA), which at z=0.678 is the most distant object in the
sample. Figure~4 shows that the moderate-redshift galaxies observed
here (11 with z$>$0.4) have a very similar size-luminosity relation to
low-redshift objects. This implies that galaxies' sizes and absolute
magnitudes have not changed very significantly since over this redshift
range, or at least that any general change in galaxy sizes has been
offset by a corresponding change in absolute magnitudes, so that the
population has shifted with redshift along, rather than perpendicular
to, the Holmberg locus.

The morphologies, colours and [OII] equivalent widths of the individual
galaxies in this sample appear to be broadly consistent with those of
various types of z$\approx$0 galaxies as they would be seen at these
redshifts. Even the three compact objects (10.2.02, 10.2.04 and
13.2.26) which have very blue colours (b$_J$$-$r$_F$=0.6--0.7) yet low
[OII] equivalent widths may have a low-redshift, low-luminosity
counterpart in the dwarf galaxy recently discovered by Steidel
\etal\ (1993).

These observations do not go deep enough to detect direct evidence of
tidal interactions in the form of distorted isophotes, trails, bridges
and other low surface brightness features. Only the presence of close
companions indicates the possibility of some interaction, and evidence
linking the companions to the star-formation must necessarily be
statistical. It is therefore of interest to compare the numbers of
close companions around the high- and low-\EW\ objects in this sample.
All five galaxies found to have close companions (10.2.11, 10.2.23,
13.2.13, 13.5.06 and 13.5.07) have \EW$>$20\AA. Thus five out of the
seventeen \EW$>$20\AA\ objects have close companions, compared to none
out of the nine \EW$<$10\AA\ objects. One possible concern about this
result is that the samples observed during the two runs had quite
different relative numbers of high- and low-\EW\ objects, and that the
use of different filters, or different exposure times or seeing, may
have made companions easier to detect in one set of data than the
other. This concern can be answered by noting that the same companions
are found on the I frames as on the B or V frames, and that the I
frames for both runs had similar integration times and reached similar
depths. Furthermore the fraction of high-\EW\ objects with close
companions is also similar for the two runs.

Is the difference in the fraction with close companions statistically
significant? If both high- and low-\EW\ galaxies had the same incidence
of close companion (\eg\ because the companions were chance
projections), the maximum likelihood of observing so many companions
about high-\EW\ objects and so few about low-\EW\ objects would be
4.5\%. Student's t-test rejects the hypothesis that the two samples are
drawn from populations with the same mean fraction of close companions
at the 1.3\% level. The conclusion that around 30\% of galaxies with
enhanced star-formation have very close ($<$10\kpc\ projected
separation) companions, whereas fewer than 10\% of low-\EW\ galaxies
have similarly close companions, is thus significant but not yet
compelling. However the numerical coincidence between the 30\% of
\EW$>$20\AA\ galaxies with close companions found here and the 40\%
excess in the number of \EW$>$20\AA\ galaxies found in the Colless
\etal\ redshift survey is striking. Although the sample is small, this
is the first direct evidence suggesting that interactions or mergers
play a significant role in the increased fraction of field galaxies
with enhanced star-formation at these redshifts.

The resolution of some objects in this notionally magnitude-limited
sample into pairs of fainter objects unfortunately makes comparisons
between the {\em absolute} fraction of close pairs in this sample and
well-studied samples at low redshift impractical. A magnitude-limited
sample of galaxies obtained at high spatial resolution is required in
order to estimate the fraction of galaxies at these redshifts in which
close companions are likely to be inducing star-formation, and to
determine whether a greater fraction of the galaxy population are
undergoing interactions than at low redshift (as suggested by Zepf \&
Koo 1989).

One cautionary point for future studies is that the colours of the
objects with close companions were in some cases misleading, since the
original photometry did not resolve the star-forming galaxy from its
(significantly redder) companion. This raises a potential difficulty in
interpreting the colour distributions of faint galaxies from photometry
with only moderate spatial resolution if merging or interactions are an
important factor.

\section{CONCLUSIONS}

A sample of galaxies from the redshift survey of Colless \etal\
(1990,1993) have been observed at high resolution using HRCam on the
CFHT. The aim was to discover whether multi-colour imaging at the best
achievable ground-based resolution could provide new information about
the physical processes behind the significant excess of star-forming
galaxies at moderate redshifts found in recent redshift surveys. The
target galaxies included a primary sample of 17 objects with [OII]
equivalent widths \EW$>$20\AA\ indicative of enhanced star-formation,
and a comparison sample of 9 objects with \EW$<$10\AA. We obtained
images in V and I or B and I with 0.5--1.0~arcsec seeing for each
galaxy, and have fitted exponential disks (or, where appropriate,
$r^{1/4}$-law bulges) directly to the 2D images in both passbands in
order to obtain objective information on the scalesizes (both relative
and absolute) of the blue and red stellar populations. We have also
searched for faint companions close to each target galaxy in order to
determine whether the star-formation might be linked to interactions or
mergers. Our main conclusions are as follows:

(i)~The galaxies observed here, covering the redshift range z=0.1--0.7,
have straightforward analogues amongst nearby objects, in that their
colours, sizes and luminosities are consistent with those of various
types of z$\approx$0 galaxies. There is no evidence for a significant
population of compact star-forming dwarf galaxies such as might be
expected if the hypothesis proposed by Babul \& Rees (1992) were solely
responsible for the excess number of blue galaxies.

(ii)~The galaxies undergoing the strongest star-formation, as indicated
by the equivalent width of [OII] 3727\AA, \EW, display a wide range of
types and absolute luminosities, from M$^\ast$$-$1 to M$^\ast$+5. In
most the blue scalelengths are comparable to or larger than the red
scalelengths, indicating global star-formation across the disk.

(iii)~Significantly, 5/17 galaxies with \EW$>$20\AA\ have fainter
companions lying within a projected distance of 10\kpc, compared to 0/9
with \EW$<$10\AA. This fraction is very similar to the fractional
excess of such galaxies observed in the redshift survey of Colless
\etal, suggesting a direct connection between interaction and
star-formation.

(iv)~Notwithstanding this evidence for interaction-related evolution,
there are several apparently isolated galaxies with large \EW\ and
three low-\EW\ galaxies with blue colours and unresolved morphologies.
Larger samples are need to determine the importance of such systems in
explaining the excess faint blue galaxy population.

\subsection*{ACKNOWLEDGEMENTS}

The observations on which this work is based were obtained at the
Canada-France-Hawaii Telescope, which is operated by the National
Research Council of Canada, the Centre National de la Recherche
Scientifique de France and the University of Hawaii. We thank the staff
of CFHT for their assistance, and Bob McClure and Mike Pierce for
sharing their expertise in the use of HRCam. The data reduction was
carried out at the Cambridge Starlink node using Starlink software and
the IRAF package.  MMC acknowledges the support of a Fellowship at
King's College, Cambridge and a Research Fellowship at the Mount
Stromlo Observatory of the Australian National University.

\end{document}